# Learning immune receptor representations with protein language models

Andreas Dounas[1,*], Tudor-Stefan Cotet[2,*], Alexander Yermanos[2,3,4,#]

[1]Sanofi, Rotkreuz, Switzerland

[2]Department of Biosystems Science and Engineering, ETH Zurich, Basel, Switzerland

[3]Institute of Microbiology, ETH Zurich, Zurich, Switzerland

[4]Center for Translational Immunology, University Medical Center Utrecht, Utrecht, Netherlands

*Equal contribution

#Correspondence: a.d.yermanos@umcutrecht.nl

**Abstract**

Protein language models (PLMs) learn contextual representations from protein sequences and are profoundly impacting various scientific disciplines spanning protein design, drug discovery, and structural predictions. One particular research area where PLMs have gained considerable attention is adaptive immune receptors, whose tremendous sequence diversity dictates the functional recognition of the adaptive immune system. The self-supervised nature underlying the training of PLMs has been recently leveraged to implement a variety of immune receptor-specific PLMs. These models have demonstrated promise in tasks such as predicting antigen-specificity and structure, computationally engineering therapeutic antibodies, and diagnostics. However, challenges including insufficient training data and considerations related to model architecture, training strategies, and data and model availability must be addressed before fully unlocking the potential of PLMs in understanding, translating, and engineering immune receptors.

**Main**

The predictive potential of adaptive immune receptor repertoires

The collection of B- and T- cells, also referred to as B- and T-cell repertoires, comprise the adaptive immune system and can recognize molecular structures (epitopes) through their highly diverse and specific B- and T-cell receptors (BCR [secreted version: antibodies], TCR, respectively). Advances in high-throughput sequencing technologies have provided a powerful approach to interrogate the immense sequence diversity underlying adaptive immune repertoires[1,2]. Millions to billions of immune receptors can be recovered within an individual study[3], thereby providing an enormous corpus of immune receptor sequences that have wide-ranging applications such as gaining fundamental insight into clonal selection, vaccine profiling, and antibody discovery.

The pervasiveness of immune repertoire sequencing across biomedical sciences has resulted in several public repositories where individual immune receptors from a variety of experimental contexts are curated, organized, and stored[4–6]. These databases differ in aspects such as size, biophysical properties, experimental contexts, and curation, all of which influence data mining and machine learning efforts. Although the number and size of repertoire repositories continue to increase, the majority of publicly available immune receptor sequences remain incomplete in terms of features such as chain pairing, structural information, and antigen specificity[7]. This has greatly complicated both the ability to look up and to predict functional properties of immune receptors from sequence alone. Nevertheless, there have been efforts to learn and infer various properties based on immune receptor sequences on both the level of the individual immune receptor and the entire repertoire[7–11]. Furthermore, open-source and community-driven frameworks have been developed that are dedicated to the intersection of machine learning and adaptive immune repertoires[9], thereby promoting reproducibility, transparency, and interpretability despite potential limitations of incomplete data.

Recent work across other disciplines has demonstrated tremendous potential in leveraging unlabeled data and unsupervised artificial intelligence for various exploratory, predictive, and generative tasks. The challenge of efficiently utilizing the implicit information within immune receptor sequences bears resemblance to those encountered in natural language processing (NLP). Both fields deal with sequential information, where context plays a crucial role in defining the significance and meaning of a sequence, be it amino acids or words. Developments in NLP such as the introduction of attention[12] and the subsequent utilization of attention mechanisms in efficient network architectures such as transformers[13] have revolutionized the field by enabling models to capture long-range dependencies and to propagate information across sequences without losing vital contextual information. Transformers take tokens as input, which refers to the smallest unit of text that is individually processed. Tokens can be a word, a word fragment or, in the case of proteins, an individual amino acid. Each token is then embedded into a specific vector in which positional information is included via positional encoding. The resulting positional input embeddings can be used for calculations in a neural network that is composed of multiple layers, with an encoder initially projecting tokens to a lower-dimensional embedding space. Encoders generate embeddings, or informative vector representations, that can then be supplied for downstream analysis or visualizations, fed into a subsequent decoder to output a new sequence of tokens, or be used as input to different neural networks with supervised learning tasks. Encoders and decoders consist of (hidden) layers, which can differ in structure between encoders and decoders but generally include a mechanism denoted as self-attention as well as feed-forward sublayers. The attention mechanism can be parallelized using multiple attention heads and allows the model to weigh how informative an input sequence's tokens are with regard to its training task. Encoders and decoders can also be used separately, and, in the context of PLMs, encoder-only models have been extensively utilized.

Such deep learning architectures for NLP, including encoders, decoders, and transformers, have been combined with self-supervised learning using massive text datasets. These models are commonly known as large language models (LLMs). During LLM training, parameters, which refer to weights (numerical values defining the strength of connections between neurons) and biases (numerical values that provide flexibility in the model's learning), are optimized based on a specific training objective, such as predicting missing or subsequent words in a sentence. Following this initial parameter optimization during self-supervised learning, the model parameters can be saved, shared and applied to new sequences that were not present in the original corpus. This original optimization of an LLM's parameters is commonly referred to as pre-training, in the sense that the model parameters are learned given an initial corpus of data and then can be used for downstream applications. Following pre-training, models can be adapted by fine-tuning using new data or a new learning objective for additional tasks. Fine-tuning a model can provide task-specific performance increases, reduced training time and data requirements, and model customization.

Alternatively, embeddings can be obtained from encoder-only models for new sequences using the originally optimized parameters. Unseen sequences can be supplied as input to a model with pre-trained weights to obtain new embeddings, which can then be utilized for further learning tasks. In doing so, the implicit information contained in the embeddings can be leveraged to boost performance in downstream tasks.

In the context of proteins, unsupervised deep learning methods employ a large corpus of potentially functional sequences and an NLP-derived pre-training task of masked language modeling (MLM) or next-word prediction (NWP) to obtain embeddings which can be further adapted to biologically relevant tasks. Such tasks can include generating viable protein designs, exploring fitness landscapes, and predicting protein structure without any explicit input evolutionary information. As sequences in protein databases are productive, generally stable, and likely functional, the learned representations of models trained on them are also assumed to be reflective of an evolutionary feasibility or "fitness". Evolutionary information can also be directly engineered into such representations by using multiple sequence alignments (MSA) as inputs and has been implemented in several models such as the MSA Transformer[14] and Evoformer[15]. Comparable to the efforts of applying NLP approaches to protein sequences, employing the unsupervised training framework for immune receptor sequences can produce embeddings that capture receptor-specific characteristics. These differ from general protein representations and are useful in interrogating immune receptors and predicting receptor-level biophysical and functional properties (Figure 1).

General PLMs
In parallel to LLMs trained on trillions of words[16], there have been similar efforts in applying language models to learn representations from large corpora of unlabeled protein sequences arising from a variety of sources. For the purposes of this review, we refer to these as general PLMs, in contrast to those language models trained specifically with immune receptor sequences. In this context, general PLMs often treat individual amino acids as tokens whereas the entire protein sequence would parallel an ensemble of words, such as a sentence[17]. Since the seminal paper describing the transformer architecture[13], there have been several iterations of general PLMs that have been further used for various sequence applications, such as protein folding[18], fitness and variant effect prediction[19], cellular localization[20], and transmembrane region annotation[21]. These general PLMs are pre-trained using large collections of raw protein sequences in a self-supervised fashion without supplying labels. Instead, the self-supervised nature of the PLMs only relies upon the sequential nature of protein tokens, aiming to either predict a set of masked residues or the next residue in a series[22], thereby learning positionally-aware patterns and contexts of amino acids at an unprecedented scale.

Examples of such general PLMs include ProtBERT[23], ESM-1b[24], ESM-2[18], Ankh[25], RITA[26], and ProtGPT2[27]. These models have been trained on the general protein databases such as BFD100, UniRef50, and UniRef100, thereby learning features from diverse collections of proteins (Table 1, Figure 2). Such protein repositories represent collections of protein sequences from many species and are of comparable size to those commonly used for text language models. For example, xTrimoPGLM[28] is a general PLM trained on both UniRef90 and ColabfoldDB[29], comprising over 1 trillion tokens and matching the dataset size of state-of-the-art text LLMs such as LLaMa[16]. However, it is important to note that not only the size of an input dataset but also its quality in terms of sequence diversity determines the efficiency of a PLM. For example, it was reported that a PLM based on UniRef50 outperformed the same models which were pre-trained on larger datasets such as UniRef100 and BFD[23,25]. Similarly, other works have explored how dataset size and model design influence the performance of general PLMs on tasks such as predicting secondary structure, contact residues, solubility, and fitness, concluding that a smaller base model (Ankh_base) could reach the same or very comparable power while requiring significantly less computational demand than a larger model (Ankh_large)[25].

The swift increase in the number of parameters has culminated in PLMs catching up to several key text LLMs, with the increase in model size corresponding to improved results. For example, xTrimoPGLM's 100 billion parameters enabled it to outperform ESM-2's 15 billion model on 12 out 15 protein fine-tuning tasks such as contact map, stability, and metal-binding prediction[28]. Furthermore, general PLMs have demonstrated promise for a wide range of problems spanning protein fold prediction[30], peptide-based drug discovery[31], synthetic protein design[32,33] and TCR specificity predictions[34,35].

Immune-receptor specific PLMs
In addition to PLMs that are pre-trained on general corpuses of proteins from various species and sources, recent works have developed PLMs trained entirely on adaptive immune receptor sequences (Table 1, Figure 2). Generally, this has involved leveraging existing transformer architectures from general PLMs but modifying the training data to exclusively include BCR and/or TCR sequences under the guise that providing immune receptors for pre-training captures immune receptor-specific context[36–38]. This data includes functional and mutational characteristics which are distinct from general proteins (e.g., due to V(D)J recombination and somatic hypermutation), thereby providing a more specific knowledge representation. Furthermore, unlike the case for general PLMs, BCR and TCR hypervariable loops are not evolutionarily conserved, thereby complicating accurate predictions in certain contexts relying upon MSAs and evolutionary constraints. In addition to variations of the underlying architectures and parameters common to general PLMs, many other immune-specific features contribute to differences in model performance such as the size and source of the pre-training data used[36,39], the nature of chain pairing[40,41], and the species included[39] (Figure 3). Furthermore, custom PLMs can be trained with receptor-specific pre-training tasks, thereby allowing the incorporation of features such as somatic hypermutation and germline ancestry (Figure 4). Receptor-specific PLMs have already demonstrated promise across a variety of tasks including *in silico* engineering of antibodies[42], humanization[39], predicting antigen-specificity[37,43] inferring receptor structure[44], designating antibody paratope sites[36], and completing missing residues[45] (Figure 5). The ability to better contextualize BCR and TCR sequences has wide-ranging implications across tasks such as structural and functional prediction, transfer learning, and biologically relevant representations of adaptive immune repertoires.

BCR-specific protein language models
BCR-specific language models have been trained to specifically contextualize functional and mutational characteristics of antibody diversity and selection. Although several models have been developed in recent years, there have been different methodologies implemented with regards to the source and curation of the training data and the underlying transformer architecture (Table 1). Furthermore, there have been various downstream applications and tasks for which a specific language model has been applied to and may be optimally suited for. It is striking to note, that although there are now many antibody-specific PLMs, these have been developed and released in only the past several years, indicating how rapidly artificial intelligence is being applied to study adaptive immune complexity. Specifically, in May 2021, directly interpretable attention mechanisms were leveraged alongside bidirectional Long Short-Term Memory (LSTM) receptor encodings to predict antibody structures using antibody sequences[46] derived from the Observed Antibody Space (OAS) database[6]. Around the same time, LSTMs had also been explored for CDRH3 generation against a specific antigen, followed by *in silico* testing of binding, developability, and classification using a model trained on experimentally validated data[47]. Transformers have surpassed recurrent neural networks as the primary architecture for NLP due to their scalability on progressively larger training datasets and due to the rise of public repositories for sharing trained models (e.g., HuggingFace[48]). Together, these factors have led to a proliferation of BCR PLMs with innovative applications.

By January 2022, multiple BCR-specific language models based on transformers were released, including AntiBERTa[36], AntiBERTy[38], AbLang[45], Sapiens[39], and IgLM[49]. These various PLMs were developed to address tasks including repertoire representation[36], predicting antibody paratope sites[36], humanization[39], predicting missing residues[45,50], and understanding B-cell evolution[38,51]. The majority of these models relied heavily upon the BERT[52] and RoBERTA[53] encoder-only architectures adapted from general PLMs, whereas IgLM adopted a decoder-only architecture inspired by GPT-2, given its goal of generative antibody design[49]. Within the specific transformer-derived architecture, there are additional computational variations involving the number of layers, the number of attention heads per layer, and the total number of model parameters that can affect a receptor-specific model's efficacy.

In addition to variations of the underlying architectures, many other variables contribute to differences in performance such as the size and source of the pre-training data used, the number of heavy and light chains, the nature of chain pairing, and the species included. For example, AntiBERTa was pre-trained on 42 million heavy chain and 15 million light chain variable regions using the RoBERTa architecture with 12 layers and 12 attention heads[36], thereby mirroring commonly sized architectures of PLMs[23,24,54]. On the other hand, Sapiens was trained on almost 40 million BCR sequences, with approximately an even split of heavy and light chains used for the pre-training data[39]. With almost 10-fold more sequences used during pre-training, AntiBERTy was created using a BERT architecture implementation hosted on HuggingFace[48] to investigate B-cell selection and evolution[38] following pre-training with 558 million unpaired natural antibody sequences from the OAS database[6]. An additional iteration of an antibody-specific PLM was developed to include biological mechanisms of B-cell evolution by tailoring two additional objectives into pre-training that learn representations of clonal relationships and somatic hypermutation based on the germline annotations from roughly 20 million BCRs in the OAS database[51].

While models such as AntiBERTa, AntiBERTy, and Sapiens were trained on heavy and light chain sequences combined into a single pre-trained model, others have been trained as chain-specific models. This includes AbLang-H and AbLang-L, which were trained using heavy and light chain sequences, respectively. Interestingly, there was a recently released model, SC-AIR-BERT, that explicitly performed pre-training on paired receptor sequences in a single model[41]. Initial investigations into feature importance and attention mapping created by SC-AIR-BERT demonstrated that paired chains during pre-training contributed to improved predictions of both BCR and TCR antigen-specificity, which, the authors argue, is in accordance with the biological importance that both chains play in structure-building and antigen-recognition[41]. This immune-receptor specific PLM is also the first to not use amino-acid level encodings but instead k-mers of size 3, as this increased vocabulary size improved performance despite increased computational costs[41].

The importance of incorporating heavy and light chain information has also been demonstrated when working with antibody embeddings generated by general PLMs. For example, embeddings created using ESM-2 and ProtT5[23] for full-length, paired sequences performed better in predicting a binary SARS-CoV-2 spike protein binding label compared to embedding the variable heavy chain sequence, the paired CDRH3-CDRL3 regions, or only the CDRH3[55]. It has further been demonstrated that training models with natively paired antibodies improved pathogen specificity classification and light chain embeddings[40]. Additional efforts to fine-tune the general PLM ESM-2 with paired and unpaired antibody sequences improved cross-chain attention on important structural regions compared to the base ESM-2 model alone, suggesting a benefit of integrating general PLMs with immune-receptor specific data to overcome the challenge and cost of obtaining paired receptor sequences[40].

Another aspect of paired chain representation learning has been translating from heavy to light chain sequences. In this context, the sequence from one chain was initially encoded and used to condition the generation of a target sequence using a ProtT5 encoder-decoder architecture[56]. For this, the 160,000 paired heavy and light chain sequences from the OAS database were considered for both forward and back translations (light-to-heavy and heavy-to-light), resulting in 320,000 paired chains in the training dataset. The fine-tuned model was able to capture V and J gene co-occurrences between input and decoded output sequence, correctly generating the target chain type in 100% of cases and V and G genes in 25% and 21%, respectively[56]. Moreover, to understand the implicit relationships between heavy and light pairs that the model was learning, attention weights between the input and target chains were plotted, indicating that hypervariable regions were less attended to and deemed less important by the model.

Although several antibody-specific PLMs have claimed the benefits of using antibody-restricted corpuses for pre-training[36,51], it is important to note that other studies have demonstrated worse performance in zero-shot (no further supervised training for a task) prediction of antibody properties such as binding affinity and melting temperature compared to general PLMs[33]. Moreover, the general PLM with the smallest number of parameters performed better than one trained on receptors on several antibody fitness-related benchmarks[33], showcasing the prevailing uncertainty regarding the best practices for pre-training receptor PLMs.

Valuable insights can also be derived from domain adaptation and how it might influence receptor-specific predictions (e.g., fine-tuning a general PLM on antibody datasets). Recent efforts towards domain adaptation of general PLMs indicate the effectiveness of this technique, as a subset of the 100 billion parameters xTrimoPGLM (xTrimoPGLM-1B) was first pre-trained on a general corpus and then fine-tuned on antibody data[28]. This produced a model exceeding the receptor-specific IgLM, AbLang, AntiBERTy, as well as general ESM2-15B, at zero-shot naturalness prediction. Naturalness here refers to a metric derived from language modeling to define how similar any given antibody is to its pre-training corpus of sequences (assumed to be "natural" repertoires). This is the opposite of perplexity, where sequences with high perplexity are not familiar to the model (lower likelihoods across all residues), whereas high naturalness and low perplexity denote a sequence relatively similar to the ones in the pre-training dataset[57]. Together, these results motivate future work to continue investigating how information from general proteins and immune receptors can be integrated to improve prediction and engineering tasks.

TCR-specific protein language models

In parallel to the progress experienced by antibody-specific PLMs, several TCR-specific PLMs have been developed leveraging various sources of publicly available data (Table 1). One of the earliest TCR-specific language models, TCR-BERT, leveraged a slightly modified BERT architecture using approximately 90,000 human and mouse TCRα and -β chains together in a single model[37]. Following pre-training on a dataset sourced predominantly from VDJdb[5] and PIRD[58], TCR-BERT was able to learn representations of TCRs from a wide range of experimental sources and antigen-specificities[37]. In contrast to the majority of BCR-specific protein language models which are trained on full-length BCRs, TCR-BERT was pre-trained entirely on the CDR3 region of both α and β chains. Another TCR-specific PLM, ProtLM.TCR, was produced using a RoBERTa style transformer model[53] trained on 62 million TCRβ CDR3 sequences originating from two sources[8,59]. Following pre-training, ProtLM.TCR was fine-tuned for the downstream task of predicting binding between TCR and HLA class I epitope sequences. Interestingly, the authors also provided additional HLA information as a categorical variable that marginally improved binding predictions (~2-4%), although it remains unclear if this increase is related to the corresponding decrease in total data size[60].

While the three aforementioned TCR-specific language models relied upon pre-training using CDR3s and incorporating epitope-specific information in subsequent steps[37,60], several of the more recent TCR-specific language models jointly embed TCR and peptide epitopes. For example, BERTrand[61], STAPLER[43], and PiTE[62] are three TCR-specific PLMs released in 2023 that explicitly model TCR-epitope relationships during pre-training, although there is variation in how they accomplish this. BERTrand and STAPLER both use a separator token to directly combine peptide:TCR sequences prior to input embedding. BERTrand only uses encoded CDR3βs with the cognate peptide, whereas STAPLER uses two separator tokens to encode CDR3α, peptide, and then CDR3β sequences, although performance using full-length and unpaired TCRα and TCRβ chains was similarly assessed[43]. Due to the difficulty in obtaining large corpuses of TCR:peptide pairs for training large language models, both BERTrand and STAPLER used strategies to generate synthetic datasets for pre-training that were subsequently fine-tuned with finite pools of annotated TCR:peptide pairs. BERTrand, for example, performed pre-training by randomly pairing peptides and TCRs using 150,000 experimentally-determined peptides from MHC-I mass spectrometry experiments and 11 million simulated TCRs using immuneSim[63]. Following pre-training, the model was fine-tuned using approximately 32,000 unique peptide:TCR pairs from various experimental sources and databases to predict antigen-specificity.

STAPLER combined known MHC-I presented peptides with TCRα and TCRβ CDR3s for initial pre-training, which involved embedding nearly 80 million random TCR:peptide combinations from 159,859 CD8 T-cell-derived sequences and 183,398 9-mer peptides that bind MHC-I. STAPLER was subsequently fine-tuned using 23,544 full-length, paired TCR sequences CD8 T-cell receptors with known binding specificity. Similar to BERTrand, STAPLER outperformed methods such as ERGO-II[64] and NetTCR-2.0[65] when predicting unseen peptide:TCR pairs[43]. As observed with SC-AIR-BERT, the core STAPLER model combining both β and α chain CDRs performed better than models only supplying a single chain, again highlighting the potential importance of pre-training with paired information.

In contrast to leveraging separator tokens and directly embedding TCR:peptide pairs, the PiTE pipeline involved separately embedding TCR and peptides to two vectors of length $l_t$ x 1024 and $l_e$ x 1024, where $l_t$ and $l_e$ are the lengths of the TCR sequence and peptide, respectively. For downstream tasks of predicting

TCR and peptide pairs, these two vectors were passed into two separate sequence encoders to obtain summarized representations that were concatenated and fed into a final classification layer. PiTE also explored a different embedding model, ELMO[66], which demonstrated the best performance on their own dataset. The authors demonstrated that the transformer-based PiTE model outperformed average pooling, BiLSTM, and CNNs when predicting antigen specificity of 290,683 unique TCRs and 982 unique epitopes derived from IEDB[67], VDJdb[5], and McPAS[68] databases.

More recently, TCR-specific PLMs have been developed to predict functional features beyond antigen-specificity that include binding affinity, epitope sequence, MHC class, and CDR3 sequences[69,70]. For example, the context-aware amino acid embedding model (catELMo) was trained on 4 million unlabeled TCR sequences from ImmunoSEQ[71] and explored how parameters such as neural network depth, architecture, and training data influenced the prediction of epitope sequences and binding affinity[69]. A common obstacle in TCR-pMHC predictions using pre-trained language models involves incomplete information pertaining to epitope, α, or β chain sequences, as well as MHC class information (e.g., only the epitope and β chain sequences could be present). To address this, an encoder-decoder model was developed (Transformer-based unsupervised language modeling for Interacting pMHC-TCR - TULIP-TCR) to decode either of the aforementioned sequences (CDR3α, CDR3β, epitope) using contextual information from the other two and from the MHC class label[70]. This approach allowed TULIP-TCR to be trained on a larger sequence corpus of 209,779 samples with at least one CDR3 and epitope annotation, supplemented by a dataset of 663,767 peptide-MHC pairs. It performed well at zero-shot affinity prediction to epitopes not present in the training data (Spearman correlation = 0.58 between predicted affinity and measured dissociation constants), and presents a potential path to more precise models for recognizing TCR epitopes.

Evaluating general and immune-receptor specific PLMs
With the recent explosion of both general and immune-receptor specific PLMs, the question arises which class of PLMs provides the best performance on immune-receptor specific tasks. General PLMs provide insight into evolutionary constraints which are not necessarily applicable to the hypervariability and V(D)J recombination mechanisms underlying the generation of adaptive immune receptors. For example, the antibody-specific PLM AntiBERTa claimed that their BCR-specific language model could better differentiate functional features of B-cells, namely naive versus memory phenotype labels, compared to the general PLM ProtBERT[36]. Similarly, a BCR-specific PLM could better predict B-cell developmental stages than general PLMs[51]. These interpretations, however, are limited to either a small number of randomly selected BCR sequences or from individual experimental sources, thereby complicating the generalizability of such findings and highlighting the need for comprehensive benchmarking focused on receptor and repertoire applications.

Multiple benchmarks have been defined to assess the performance of general PLMs in the context of immune receptors, consisting of either deep mutational scanning (DMS) datasets with a known receptor property (e.g., affinity, expression, developability) for zero-shot fitness predictions or specialized datasets for model fine-tuning (e.g., antibody structure prediction). For example, the general PLM ProGen2 performed zero-shot fitness prediction assessment on deep mutational scanning data with measured antibody binding affinities, verifying if the higher likelihood PLM mutation corresponded to a better affinity in the experimental set[33]. Similarly, the pAbT5 BCR chain translation model[56] was evaluated by correlating zero-shot perplexity with several fitness scores (stability, binding affinity, expression), achieving a performance similar to ProGen2 and better than a model trained using BCRs (ProGen2-OAS). The subset of the xTrimoPGLM model trained on OAS, xTrimoPGLM-Ab-1B, was used to predict the naturalness of a given antibody in the same zero-shot setting and outperformed several BCR-specific models such as IgLM and AntiBERTy[28]. Lastly, the Solvent framework[72] scrutinized the use of receptor-specific (AntiBERTy) versus general (ESM-2) language model embeddings for antibody structure prediction. This revealed that ESM-2 trained on 650 million sequences could improve CDRH3 loop prediction, whereas the smaller ESM-2 (35 million sequences) performed similarly to AntiBERTy across all CDRs.

Although the aforementioned studies explicitly compare model performance on specific tasks, it remains unclear how to robustly assess and evaluate the performance and representation of PLMs across various tasks specific to adaptive immune receptors. To address this challenge, a study benchmark, the **AnTibody Understanding Evaluation (ATUE)**, was recently introduced to evaluate and compare the performance of general PLMs compared to BCR-specific PLMs[51]. The authors developed a BCR-specific language model that explicitly incorporated evolutionary features, and demonstrated that BCR-specific PLMs performed better than foundation PLMs on antibody-specific tasks across varying degrees of specificity. These tasks included antigen-binding classification, paratope prediction sequence labeling, B-cell maturation stage classification, and antibody discovery classification. While general and antibody-specific PLMs performed similarly on those tasks less specific to antibodies, namely antigen binding and paratope prediction, antibody-specific PLMs performed better on tasks involving B-cell maturation stage classification and antibody discovery classification[51]. Although the introduction of benchmarks to assess the representation and evaluation of language models represents an important progression in directly comparing the performance of PLMs, it should be noted that the scarcity of high-quality immune receptor data limits the ability to compare the performance of PLMs on antibody-specific tasks.

Another publication compared the performance of general (ESM-2) and receptor-specific PLMs (AbLang and TCR-BERT) on several receptor and repertoire-specific applications[73]. These ranged from receptor binding predictions for a given epitope, correlating embedding and Levenshtein distances in different hidden layers, correct V-, J-gene, clonotype assignment, and clonal tree inference in the embedding space. The work highlighted the nuanced nature of evaluating PLMs, as many features such as model size and selection of layers impacted model performance. For example, training using embeddings from earlier layers of ESM-2 models yielded significantly better performance for several metrics compared to models trained on embeddings from the last layer, whereas this was not observed for a TCR PLM[73]. Additionally, embeddings produced from general PLMs demonstrated a higher correlation to sequence distances than an antibody-specific PLM, thereby giving insight into how language models trained on large corpuses can generalize elementary concepts such as sequence similarity.

As PLMs have been proven effective in zero-shot fitness prediction, the FLAb (**F**itness **L**andscape for **A**nti**b**odies) benchmark was introduced to assess the performance of both general and receptor-specific PLMs for 6 fitness proxies in the context of therapeutic antibody design (expression, thermostability, immunogenicity, aggregation, polyreactivity, binding affinity)[74]. This was based upon the concept that a functional PLM should capture biophysical properties of an antibody, and therefore, a high-fitness antibody should be associated with a low perplexity score. This was applied to several general and receptor-specific PLMs to reveal that ProGen2-small performed the best on 7 out of the 20 available datasets. Moreover, minor differences were observed when contrasting models with different architectures trained on the same dataset, models with varying number of parameters (e.g., ProGen2-small, ProGen2-medium, ProGen2-large, ProGen2-xlarge), or structure-conditioned models versus sequence-only ones. Overall, this work introduced an essential set of benchmarks to guide the architecture and training dataset decision-making when developing new PLM models for designing antibody therapeutics.

In contrast to deciding between general and immune-receptor specific PLMs, a recent work presented a transfer learning approach to combine the strengths of both modalities. The proposed AbMAP (Antibody Mutagenesis-Augmented Processing) pipeline first leverages general PLMs but then incorporates antibody-specific information to learn representations of the hypervariable regions. Unlike most other BCR-specific PLMs developed, AbMAP focuses entirely on the hypervariable region and four flanking framework residues given this region's importance in determining antibody function. The authors argue to accentuate the CDR-specific context by generating novel sequences via *in silico* mutagenesis and comparing their embeddings from those of originally recovered CDRs. This transfer learning approach was applied to several tasks associated with varying degrees of antibody-specificity, such as identifying structural templates, predicting the binding energy changes from mutations, and paratope identification. AbMAP was able to outperform state-of-the-art structural prediction techniques such as AlphaFold2 and DeepAb, and performed comparably to general PLMs on paratope identification. Overall, transfer learning approaches such as AbMAP provide example frameworks to rapidly adapt novel general PLMs by fine-tuning with receptor-specific information.

Incorporating structural information into PLMs
Recent involvements in protein language modeling have been centered around supplying information from other modalities such as descriptive text[75] or structures[76–78] into the sequence embeddings. For example, AbMAP employs an approach to leverage immune receptor structural information[79]. To obtain a more informative representation of the CDR loop regions, AbMAP maximizes the information captured in CDR embeddings by enforcing them to learn structural and functional similarities in a multi-task learning setting. For this, AbMAP feeds the general PLM embeddings of paired sequences into a transformer encoder module, predicting both the TM-score (structural similarity) and binding specificity score (functional similarity). The resulting model, AbMAP-B, outperformed AlphaFold2 and OmegaFold after fine-tuning for antibody structure prediction, indicating that a supervised refinement approach that leverages receptor structures is better suited in overall CDR loop modeling. This strategy of incorporating sequence and structural features of immune receptors has similarly been leveraged by DeepAIR, which also integrated transformers and AlphaFold2[15] structures to predict binding affinity, reactivity, and immune repertoire classification[80].

Other efforts have incorporated structural information by training a transformer to map protein backbone structures to a set of sequencers, which is referred to as inverse folding[81]. For example, a recently released inverse folding model for antibodies, AntiFold, was developed to improve antibody structure design and optimization[82]. The inverse folding architecture underlying this model, ESM-IF1, has previously been leveraged for zero-shot binding affinity prediction between the SARS-CoV-2 receptor binding domain and ACE2[83]. Unlike ESM-IF1, AntiFold was trained on antibody structures to account for the distinct structural and sequence properties inherent to antibodies. Overall, AntiFold had a notably higher sequence recovery rate of all CDR regions compared to general inverse folding models and an antibody-adapted version of ProteinMPNN[84], and could further be used to generate antibodies with high structural similarity to known experimental structures[82]. Another approach, IgDesign, has leveraged inverse folding to design antibodies *in silico* and subsequently validated binding with experimental assays[85]. Together these recent works emphasize the promise of inverse folding to generate novel sequences conditioned on structural features.

Inspired by the recent image generative models prompted by text, structural information has similarly been integrated into PLM's representations using a contrastive learning approach. Specifically, AntiBERTa-2 implemented Contrastive Language-Image Pre-training[86], which aims to maximize the cosine similarity between the embeddings of a given correct text-image pair (or sequence-structure pair in this case) and minimize the similarity between incorrect pairs, to refine AntiBERTa's sequence embeddings with structural information. To extract structural information, the frozen ESM-IF1 inverse folding model's encoder was used during training. In total, 1,237 human antibody structures obtained from SAbDab[87] were used for contrastive pre-training and produced a model that better correlates with antibody structural similarity compared to the sequence-only AntiBERTa[36] and AntiBERTy[38] for both CDRH3 and variable regions. It was observed that supplying computationally predicted protein structures or combining predicted and crystal structures did not provide novel information to their learning framework. This consideration holds important ramifications for training future receptor-specific PLMs given the sparsity of experimentally solved structures.

Learning repertoire-wide representations with PLMs
In addition to the aforementioned tasks such as antigen-specificity prediction, paratope identification, and humanization, there have been several examples of leveraging PLMs to quantify selection and diversity of immune receptors. One of the most common methods involves leveraging embeddings from the hidden layers of PLMs. For example, the output embeddings from the last hidden layer of antibody-specific and general PLMs were used to visualize how embeddings could quantify features of adaptive immune repertoires including germline gene usage, somatic hypermutation, and B-cell phenotype[36]. While in many models the embeddings derived from the last layer are utilized, this is neither necessary nor guaranteed to be optimal as relevant tokens for properties of interest may receive more attention in intermediate layers[73,88]. Performing uniform manifold approximation projections on the averaged embeddings from the last hidden layer demonstrated clear clustering based on V-gene for all three models, whereas only projections using AntiBERTa revealed distinct separation between naive and memory B-cell populations[36]. This is in accordance with the antibody-specific PLM AbLang, whose embeddings could better resolve naive and memory B-cells when compared to the general PLM ESM-1b[45]. Embeddings of BCRs have also demonstrated the ability to capture features of somatic hypermutation[36,45,51], which have been further related to evolutionary divergence from germline sequences and how B-cell evolution relates to PLM-based fitness scores[38]. In the realm of T-cell repertoire selection, the comparative analysis between general and TCR-specific PLMs remains somewhat underexplored. However, the visualization of TCR-BERT embeddings for human TRB sequences hints at specific clusterings linked to antigen-specificity[37].

These examples of applying PLMs to describe features of adaptive immune repertoires (e.g., germline gene usage, somatic hypermutation, and cell phenotype) were largely exploratory investigations into how PLMs could describe subsets of repertoires[36,45] or individual clonal lineages[38,51]. Recently, the transfer learning approach AbMAP was used to investigate how BCRs from different individuals are distributed across PLM embedding spaces and whether PLMs could help identify convergent immune signatures[79] given recent reports of the tremendously personalized nature of immune repertoires[3]. Indeed, AbMAP detected convergent signatures using their PLM-based framework that may not immediately be clear when assessing only sequence similarity[79]. The authors further argue that these high-occupancy regions of PLM embedding space may be useful for the development of therapeutic antibody drugs.

PLM-derived learned representations of immune repertoires could additionally be integrated into classifiers for patient-level repertoire diagnostics. For example, a recent method, graph representation of immune repertoires (GRIP), represented entire immune receptors as embedding vectors using graph neural networks and transformers and investigated the potential of predicting survival probabilities of cancer patients[89]. By creating a sequence-similarity graph using ProtBERT to extract node features, and subsequently feeding this into graph neural networks (GNNs) with graph pooling, GRIP extracted subclone-level representations, which were later converted into a repertoire embedding (via a GNN and transformer architecture) for predicting patient outcomes. Interestingly, GRIP's architecture enables an interpretable analysis of the model's predictions via integrated gradients, indicating the risk level of a single sequence node. In the context of repertoire diagnostics, Mal-ID (MAchine Learning for Immunological Diagnosis) was recently reported to leverage the feature-rich landscape of B and T-cell repertoires to accurately predict immune states ranging from SARS-CoV-2, HIV, systemic lupus erythematosus, common variable immunodeficiency, and healthy controls[10]. Although repertoire-wide diagnostics have been explored in the past[8,11,90], this recent platform integrated multiple modalities including a lasso linear model, signatures of convergent clones, and PLMs to achieve high accuracy classification[10]. The authors appended the three heavy chain CDR segments of each receptor and explored embedded sequences using either the default UniRep general weights or fine-tuned using antibody-specific information. Although this framework was only applied to five distinct immune states, it nevertheless demonstrates the applicability of PLMs to repertoire-wide readouts of health and disease status.

Further applications of PLMs in adaptive immunity
Beyond learning repertoire-scale representations, PLMs have demonstrated promise in other areas of adaptive immunity, including domains such as structural prediction of immune receptors[72] and therapeutic optimization[91]. ESM-Fold, for example, leverages a large language model of 15 billion parameters to accurately and rapidly predict protein structures[18]. Recently, IgFold[38] was applied to predict structures of paired antibody sequences with comparable or higher accuracy than other popular methods[44] such as RepertoireBuilder[92], DeepAb[46], ABlooper[93], and AlphaFold-Multimer[94], although may be less performant than other models such as AlphaFold-Multimer, XtrimoPGLM[28] and XtrimoABFold[95]. IgFold, uses AntiBERTy to convert antibody sequences into embeddings and subsequently uses a series of graph transformer and invariant point attention layers to predict atomic coordinates for the protein backbone atoms, with a final refinement step using Rosetta[96]. While AlphaFold-Multimer and ESM-Fold have incorporated the ability to simultaneously model antibodies and antigen interactions, IgFold is restricted to only antibody sequences. It therefore remains unknown how future iterations of protein-language models pre-trained using receptors and antigens can improve the structural reconstruction and prediction of biophysical features such as specificity, affinity, and neutralization. A particular use of antibody folding models is hallucination for iteratively designing an antibody sequence to match a given structure. For example, a library of Trastuzumab CDRH3 variants was generated using the LSTM-based DeepAb and then screened for HER2 interaction using Rosetta, akin to an *in silico* DMS experiment[97]. Their library was compared to an experimentally derived library[98] and demonstrated congruent results between the experimental and synthetic libraries.

PLMs have also been suggested as a computational alternative to the expensive and laborious experimental procedure of improving antibody developability. Specifically, recent work has leveraged PLMs to improve the thermostability of single-chain variable fragments[91], as this is often a hurdle for therapeutic antibody implementation. Using temperature-specific data from multiple scFv libraries, the authors noted that a simple CNN model performed better than general PLMs, whereas there was an advantage in using an antibody-specific PLM in classifying thermostable scFv variants and identifying important residues[91].

Recent work has also leveraged general PLMs for the *in silico* affinity maturation of monoclonal antibodies[42]. Antibody affinity was improved for seven antibodies by screening 20 or fewer variants by leveraging the evolutionary rules computed by general PLMs. Subsequent inverse folding approaches were similarly used to guide evolution for improving the affinity and neutralization of clinical antibodies used to treat SARS-CoV-2 infection[99]. Finally, PLM embeddings have been employed for the *in silico* exploration of the sequence space of high-affinity antibodies[32]. For example, an ensemble of affinity regression models was fine-tuned from BERT models using known binding affinities. These affinity predictors were further employed in exploring the fitness landscape of seed sequences, where residues were picked and mutated according to various sampling strategies and mutations if they improved the predicted affinity. This resulted in a designed library with diverse sequences and a 28.7-fold affinity improvement for the best binder compared to a standard directed evolution library[32]. Together, these approaches exemplify that using pre-trained general PLMs and fine-tuned models on affinity data can generate potent fitness predictors that can be used for exploring fitness landscapes and engineering antibody therapeutics.

Challenges and future perspectives
Learning the language of the adaptive immune system has already unleashed unquantifiable benefits for human health, with applications ranging from therapeutic antibody treatments, vaccination design, personalized medicine, and immunotherapy. Given the sequence-rich nature underlying adaptive immune recognition, it is conceivable that approaches guided by NLP can accelerate our ability to understand, translate and engineer immune receptors. Here, we have outlined some of the most recent developments of how PLMs are being applied to adaptive immunity. While these early reports highlight the benefit of applying PLMs to adaptive immune receptors, there remain several obstacles before the full potential of PLMs can be unleashed. Many challenges relate to the insufficient amount and quality of training data, created by an experimental bottleneck, for determining important properties such as antigen specificity, affinity and structure. Although the fast-moving field of AI will continue to produce improved computational algorithms, these will remain limited in their applicability to repertoires without investment into innovative strategies to generate the necessary training data. Coordinated efforts to curate, annotate, and share data used for training and fine-tuning will become increasingly valuable in assessing and benchmarking the performance of newly developed PLMs.

Many open questions remain regarding how the nature of the immune receptor corpus influences learned representations and model performance. For example, factors such as whether pre-training should be performed on general proteins or exclusively on immune receptors, full-length sequences versus CDR3s, receptor-antigen interactions are included, or multiple species or even individuals will influence downstream conclusions and predictions. All of these considerations are even prior to the selection of the AI model (e.g., BERT, GPT, RoBERTa), complexity (e.g., number of layers, attention heads, and parameters), how the embeddings are processed (average of last hidden or intermediate layer versus only the CLS token), and tokenization (e.g., amino acids versus k-mers). The exploration and optimization of these factors similarly imply a financial limitation given the extensive computational demands sometimes required for sufficient training.

Future work can be conducted towards learning informative representations of entire immune repertoires. Effectively understanding repertoire embeddings may help to identify features that are indicative of shared biological underpinnings ranging from immune exposure to disease status. Defining metrics and frameworks to quantify and predict selective features of immune repertoires can advance our fundamental understanding of immunology and can help computationally optimize, design and discover therapeutics. Contrasting receptor-specific and general protein language models and investigating the effect of infusing receptor structural information at the embedding level are also fundamental areas that should be explored in the context of learning immune repertoire representations using PLMs.

Beyond technical considerations, there remain concerns regarding the safety, ethics, biases, and equity of how PLMs are applied to adaptive immune receptors. Questions over the property and privacy of training data will become increasingly relevant, as an individual's immune receptors will be used to design and engineer novel therapeutics. Furthermore, biases with respect to which geographic, ethnic, sex, and age populations are underrepresented or omitted in the corpuses used to train or fine-tune language models need to be seriously considered and managed as novel language models continue to be developed. The recent movement to increase the interpretability of AI is motivated by the need to uncover these specific biases that are hidden behind increasingly complex models. Finally, principles promoting equitable and transparent AI should be prioritized in the context of PLMs. For example, while several receptor-specific PLMs publish the training data, the pre-trained model weights, and underlying code, this is not always the case. Establishing and adhering to a standard data and model sharing is essential to allow equitable engagement with the ongoing AI revolution in adaptive immunity.

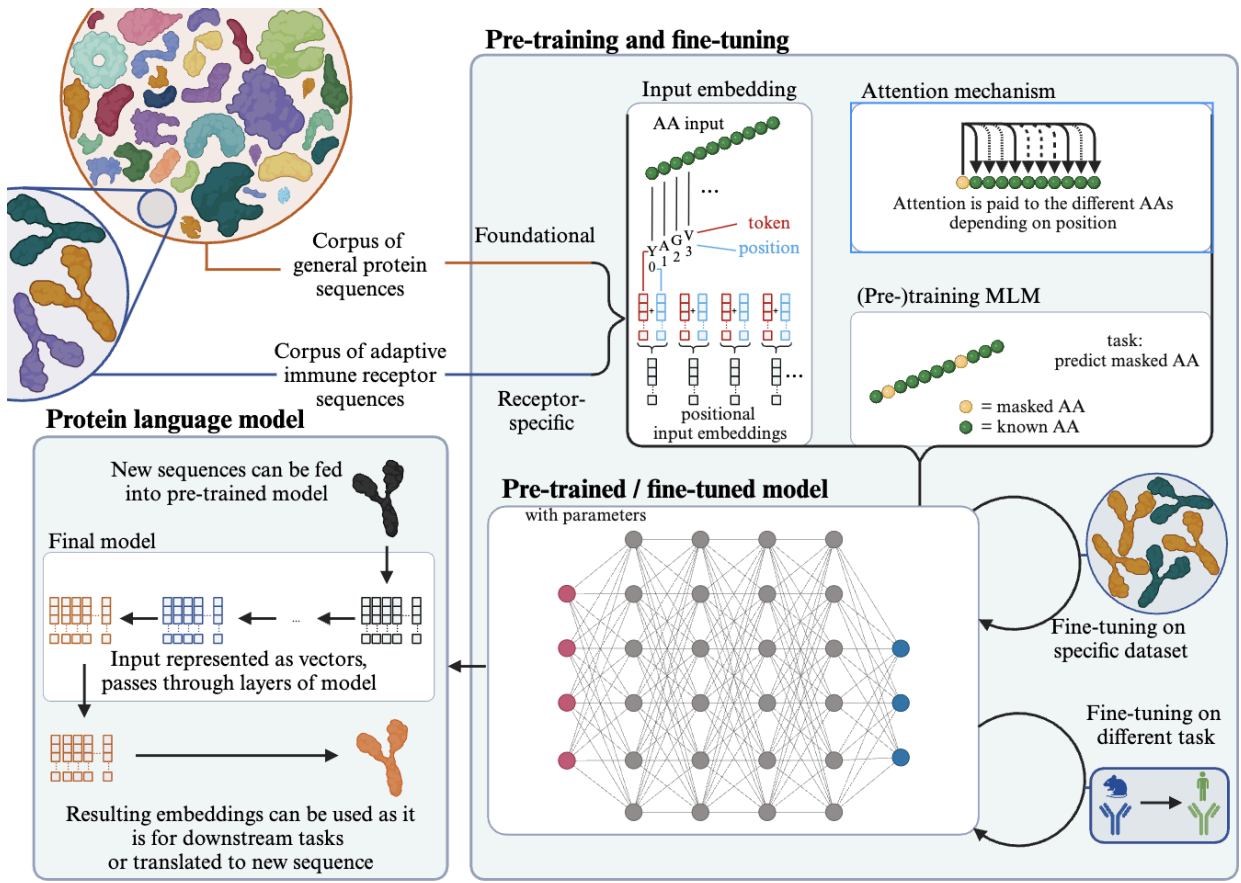

**Figure 1. Graphical overview of general and receptor-specific protein language models (PLMs).** PLMs have been trained using either protein sequences from diverse corpuses of general proteins from various species and protein families (general) or immune-receptor specific. Following pre-training and fine-tuning, the model parameters can be saved, shared and applied to new sequences that were not present in the original corpus.

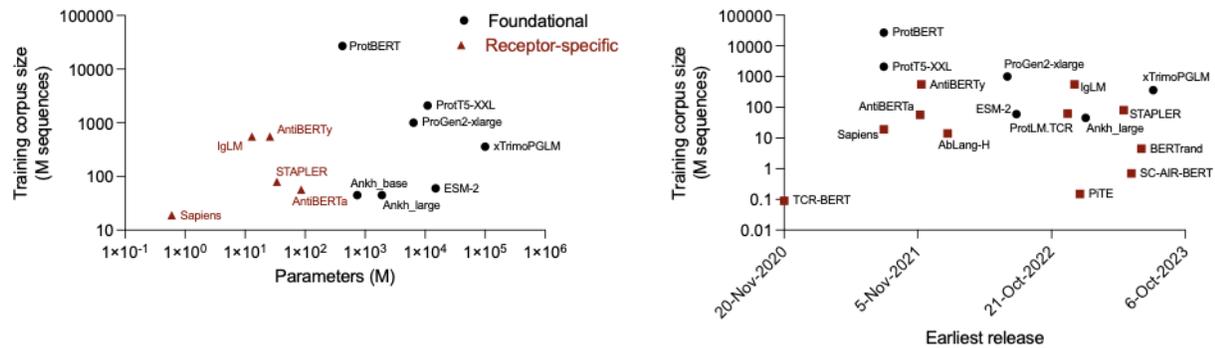

**Figure 2. Timeline of select PLM release, corpus size, and model parameters.** General (models trained on diverse protein families and species) and immune-receptor specific PLMs have been trained using millions (M) of sequences and parameters over recent years. The largest value was selected in cases where multiple models using the same architecture and name were trained.

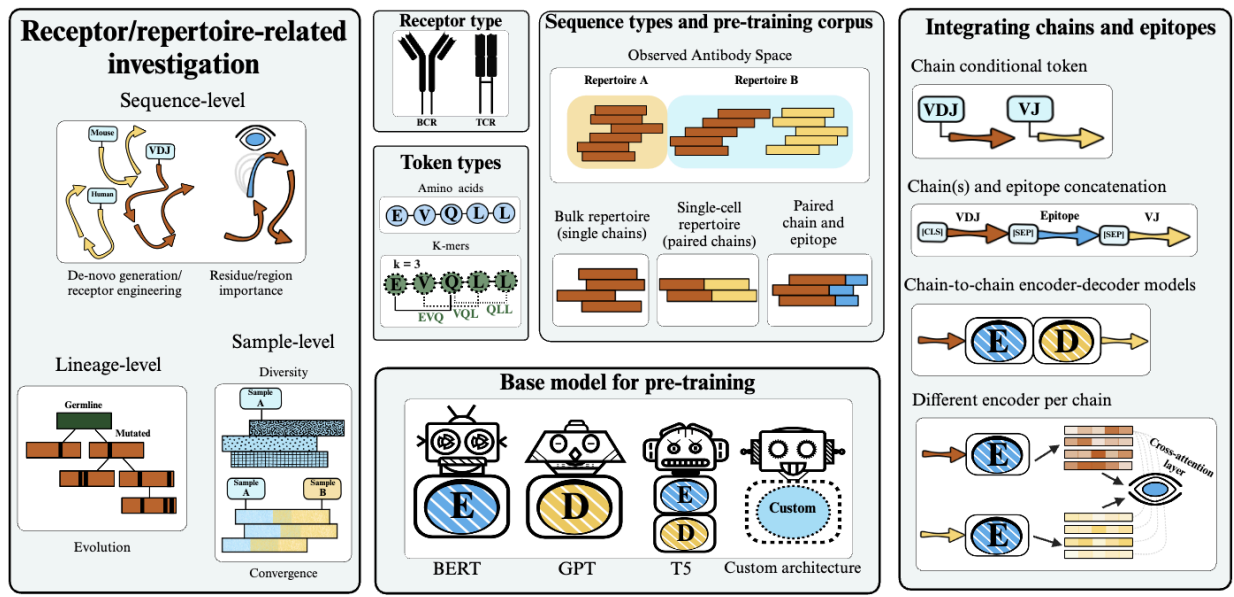

**Figure 3. Immune-specific considerations for pre-training PLMs.** A variety of factors including type of receptor (species, chain), AI model architecture (Encode (E), Decode (D)), architecture size (number of layers, hidden layer size), and incorporation of epitope sequence influence output embeddings and model performance.

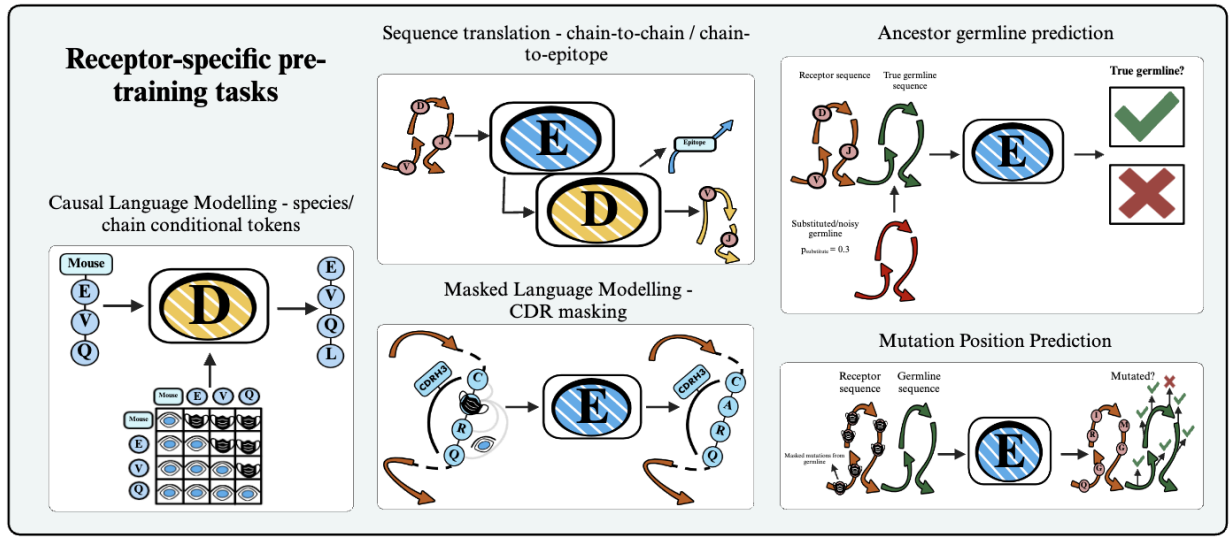

Figure 4. Receptor-specific pre-training tasks can be implemented to incorporate immune-specific properties such as ancestor germline and somatic hypermutation.

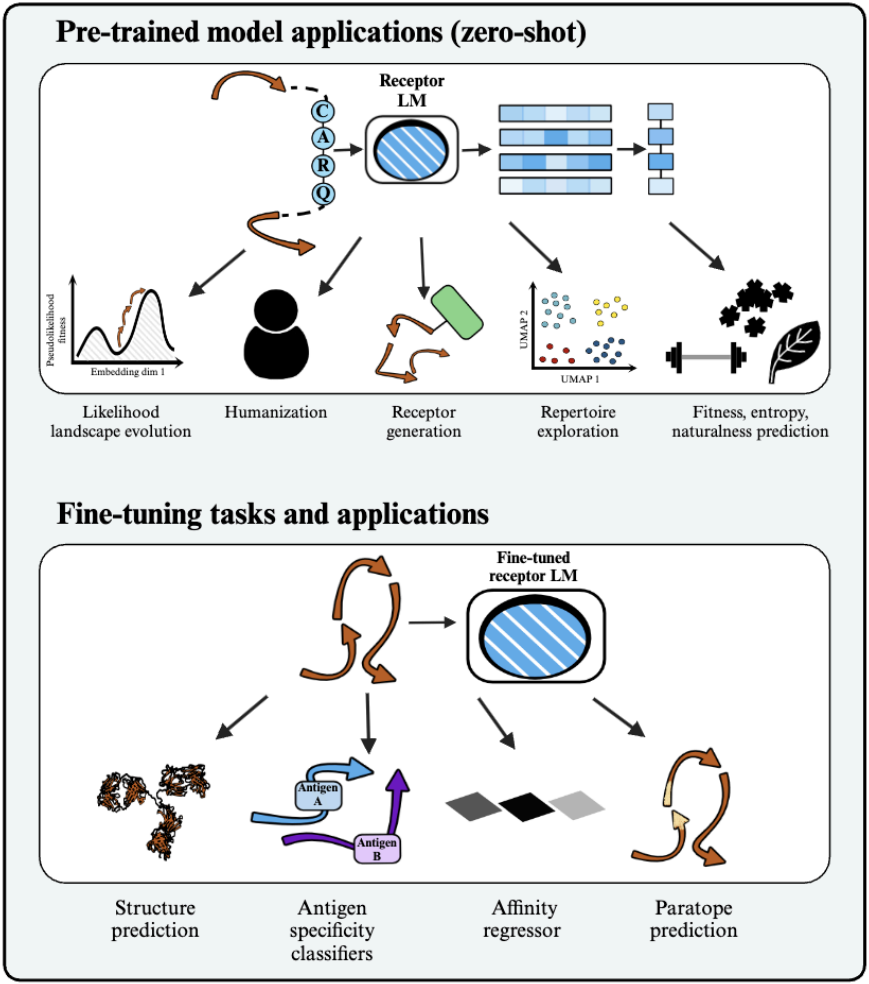

Figure 5. Examples of downstream applications of pre-trained and fine-tuned protein language models.

| name | layers | heads | embed dim | ffl dim | params (M) | corpus size (M) | database | sequence |
|---|---|---|---|---|---|---|---|---|
| ProtBERT | 30 | 16 | 1024 | 4096 | 420 | 27000 | BFD100 / UniRef100 | protein |
| ESM1b | 33 | 20 | 1280 | 5120 | 650 | 250 ? | UR50/S | protein |
| ESM-2* | 48 | 40 | 5120 | 20480 | 15000 | 60 | UR50 (sample UR90) | protein |
| ProtT5-XL | 24 | 32 | 1024 | 16384 | 3000 | 5000 | UniRef50 / BFD100 | protein |
| ProtT5-XXL | 24 | 128 | 1024 | 65536 | 11000 | 2100 | UniRef50 / BFD100 | protein |
| Rita_small | 12 | 12 | 768 | 3072 | 125 | 150 AAs | UniRef100, Metaclust, MGnify | protein |
| Rita_medium | 24 | 16 | 1024 | 4096 | 350 | 150 AAs | UniRef100, Metaclust, MGnify | protein |
| Rita_large | 24 | 16 | 1536 | 6144 | 760 | 150 AAs | UniRef100, Metaclust, MGnify | protein |
| Rita_XL | 24 | 24 | 2048 | 8192 | 1200 | 150 AAs | UniRef100, Metaclust, MGnify | protein |
| ProGen2-small | 12 | 16 | 1024 | 4096 | 151 | 1000 | UniRef90 /BFD30 | protein |
| ProGen2-medium | 27 | 16 | 1024 | 4096 | 764 | 1000 | UniRef90 /BFD30 | protein |
| ProGen2-large | 32 | 32 | 1024 | 4096 | 2700 | 1000 | UniRef90 /BFD30 | protein |
| ProGen2-xlarge | 32 | 16 | 1024 | 4096 | 6400 | 1000 | UniRef90 /BFD30 | protein |
| Ankh_base | 48+24** | 12 | 768 | 3072 | 740 | ? | UniRef50 | protein |
| Ankh_large | 48+24** | 16 | 1536 | 3840 | 1900 | ? | UniRef50 | protein |
| xTrimoPGLM | 72 | 80 | 10240 | 31744 | 100000 | 360 | UniRef90 /ColAbFoldDB | protein |
| AntiBERTa | 12 | 12 | 768 | 3072 | 86 | 42 / 15 | OAS | BCR LH (u) |
| AntiBERTy | 8 | 8 | 512 | 2048 | 26 | 558 | OAS | BCR LH (p) |
| AbLang-H | 12 | 12 | 768 | 3072 | ? | 14 | OAS | BCR H |
| AbLang-L | 12 | 12 | 768 | 3072 | ? | 0.2 | OAS | BCR L |
| IgLM | 4 | 8 | 512 | 2048 | 13 | 558 | OAS | BCR LH (p) |
| IgLM-S | 3 | 6 | 192 | 786 | 1.5 | 558 | OAS | BCR LH (p) |
| Sapiens | 4 | 8 | 128 | 256 | 0.6 | 10 | OAS | BCR H |
| Sapiens | 4 | 8 | 128 | 256 | 0.6 | 19 | OAS | BCR L |
| SC-AIR-BERT | 6 | 4 | 512 | 2048 | ? | 0.85 / 0.7 | multiple | TCR$\alpha\beta$ & BCR LH CDR3 (p) |
| TCR-BERT | 12 | 12 | 768 | 1536 | ? | 0.09 | VDJdb, PIRD / PIRD | TCR$\alpha\beta$ |
| STAPLER | 2 | 2 | 128 | 512 | 0.55 | 80 | multiple | TCR CDR3$\alpha\beta$ |
| STAPLER | 4 | 4 | 256 | 512 | 4.3 | 80 | multiple | TCR CDR3$\alpha\beta$ |
| STAPLER | 8 | 8 | 512 | 512 | 33.9 | 80 | multiple | TCR CDR3$\alpha\beta$ |
| STAPLER | 2 | 2 | 128 | 128 | 0.55 | 80 | multiple | TCR$\alpha\beta$ |
| STAPLER | 4 | 4 | 256 | 128 | 4.3 | 80 | multiple | TCR$\alpha\beta$ |
| STAPLER | 8 | 8 | 512 | 128 | 33.9 | 80 | multiple | TCR$\alpha\beta$ |
| PiTE | 1 | 2 | 1024 | 32 | 20 | 0.3*** | IEDB, VDJdb, McPAS | TCR CDR3$\beta$+epi |
| BERTrand | 12 | 12 | 768 | 3072 | ? | 4.5 | immuneSIM+Abelin, Di Marco, Faridi, Sarkizova | TCR CDR3$\beta$+epi |

**Table 1. Architecture hyperparameters of select protein language models, the size of pre-training corpus ("corpus"), and the type of utilized input data ("sequence").** *only one of multiple models mentioned. **48 encoder layers and 24 decoder layers. ***of which half are binding and half are

non-binding. embed dim: embedding dimension. ffl dim: feedforward layer dimension. AA: amino acid. p: paired. u: unpaired. BCR: B cell receptor. TCR: T cell receptor. L: light chain. H: heavy chain. α: alpha chain. β: beta chain. CDR3: complementarity determining region 3. epi: epitope.

| name | git |
|---|---|
| AntiBERTa | https://github.com/alchemab/antiberta |
| AntiBERTy | https://github.com/dohlee/antiberty-pytorch |
| AbLang-H | https://github.com/oxpig/AbLang |
| IgLM | https://github.com/Graylab/IgLM |
| Sapiens | https://github.com/Merck/BioPhi-2021-publication |
| ESM-2 | https://github.com/facebookresearch/esm |
| ESM1b | https://github.com/facebookresearch/esm |
| TCR-BERT | https://github.com/wukevin/tcr-bert |
| STAPLER | https://github.com/NKI-AI/STAPLER |
| PiTE | https://github.com/Lee-CBG/PiTE |
| BERTrand | https://github.com/SFGLab/bertrand |
| SC-AIR-BERT | https://github.com/TencentAILabHealthcare/SC-AIR-BERT |

**Table 2. Git repositories of select protein language models.**